\documentclass[twocolumn,eqsecnum,superscriptaddress,showpacs]{revtex4}
\usepackage{bm}

\begin{document}
\title{ New Concept of First Quantization}
\author{Takehisa \textsc{Fujita}} \email{fffujita@phys.cst.nihon-u.ac.jp}
\author{Seiji \textsc{Kanemaki}} \email{kanemaki@phys.cst.nihon-u.ac.jp}
\affiliation{Department of Physics, Faculty of Science and Technology
Nihon University, Tokyo, Japan }%
\author{Sachiko \textsc{Oshima}} 
\affiliation{Department of Physics, Faculty of Science, Tokyo Institute 
of Technology, Tokyo, Japan} 

\date{\today}

\begin{abstract}

We derive a Lagrangian density of Dirac field by employing the local gauge invariance 
and the Maxwell equation as the fundamental principle. The only assumption made here 
is that the fermion field should have four components. The present derivation of the Dirac 
equation does not involve the first quantization, and therefore this study may present 
an alternative way of understanding the quantization procedure. 

\end{abstract}

\pacs{11.15.-q,03.50.-z,03.50.De,03.65.-w}

\maketitle

\section{ Introduction}
Dirac derived the Dirac equation by factorizing Einstein's dispersion relation 
such that  the field equation becomes the first order in time derivative \cite{dirac}. 
Namely, he factorized the relativistic dispersion relation employing four by four matrices
$$ E^2 -{\bm{p}^2-m^2 }=
(E-\bm{p}\cdot \bm{\alpha}-m\beta )
(E+\bm{p}\cdot \bm{\alpha}+m\beta )  \eqno{(1.1)}  $$
where $\bm{\alpha}$ and $\beta$ are Dirac matrices which satisfy the following 
anti-commutation relations
$$ \{ \alpha_i, \alpha_j \} =2\delta_{ij}, \ \ \ \ \ \{ \alpha_i, \beta \} =0, 
\ \ \ \  \ \  \beta^2=1. $$
Here, $i$ and $j$ run $i,j=x,y,z$. 
In this case, Dirac equation becomes 
$$ \left(i\hbar {\partial\over{\partial t}}+i\hbar \bm{\nabla}\cdot \bm{\alpha}
-m\beta \right)\psi=0  \eqno{(1.2)}  $$
where $\psi$ denotes the wave function and should have four components since 
$\bm{\alpha}$ and $\beta$ are four by four matrices, 
$$ \psi = \pmatrix{ \psi_1 \cr 
 \psi_2 \cr 
\psi_3 \cr 
\psi_4 \cr }. \eqno{(1.3)} $$                   
In eq.(1.2), the first quantization condition
$$ E\rightarrow i\hbar {\partial\over{\partial t}}, \ \ \ \ \ \ 
\bm{p}\rightarrow -i\hbar \bm{\nabla} \eqno{(1.4)} $$
is employed. Even though the first quantization is not  
understood very well from the fundamental principle, the Dirac equation and the first 
quantization are, of course, consistent with experiment.  
Nevertheless, it should be interesting if one can derive the Dirac equation 
without involving the first quantization condition. 

In this Letter, we derive the Lagrangian density of the fermion fields 
interacting with the gauge field ${A}^\mu$. This is done by employing 
the Maxwell equation as the most fundamental equation and by requiring 
the local gauge invariance of the Lagrangian density, but the first quantization condition 
of eq.(1.4) is not employed at all. The basic assumption we have made, 
in addition to the local gauge invariance, is that the fermion field 
should have four components and the Lagrangian density should be 
written by Lorentz scalars. 

In the present picture, we can immediately obtain two interesting results. 
The first consequence from the new picture is that the Klein-Gordon equation 
has lost its foundation. This can be easily seen, since the first quantization condition 
is not the fundamental principle any more, it cannot be used for the replacement of 
the energy and momentum by the differential operators in the squared relativistic 
dispersion relation. The second result is connected to the well-known wrong Hamiltonian 
which is obtained by the canonical quantization procedure in polar coordinates. 
This is now understood well since the first quantization is obtained from the Dirac 
equation as the consequence by the identification of the momenta in terms 
of the differential operators. 

\section{Lagrangian Density for Maxwell Equation}
Now, we discuss a new way of deriving the Lagrangian density 
of Dirac field by employing the local gauge invariance and the Maxwell equation 
as the starting point. Since the Maxwell equation is a field equation, 
the first quantization is already done there. Therefore, we consider that the Maxwell 
equation should give a guide to constructing the Lagrangian density 
for fermion fields. 

First, we start from the Lagrangian density that reproduces the Maxwell equation
$$  {\cal L} = - gj_\mu A^\mu
  -{1\over 4} F_{\mu\nu} F^{\mu\nu}   \eqno{ (2.1)} $$
where $A^\mu$ is the gauge field, and $F_{\mu\nu}$ is the field strength 
which is defined as
$$   F_{\mu\nu}= \partial_{\mu} A_{\nu} - \partial_{\nu} A_{\mu} . $$
$j_\mu$ denotes the current density of matter field which couples 
with the electromagentic field. From the Lagrange equation, one obtains 
$$ \partial_{\mu} F^{\mu\nu} = g j^{\nu}  \eqno{ (2.2)} $$
which is just the Maxwell equation. For the case of no external current ($j^{\mu}=0$), 
we obtain the equation for the gauge field $\bm{A}$ 
$$ \left({\partial^2\over{\partial t^2}} -\bm{\nabla}^2 \right)\bm{A}=0 \eqno{ (2.3)} $$
where we have made use of the Coulomb gauge fixing 
$$ \bm{\nabla}\cdot \bm{A} =0 . $$
It is clear that eq.(2.3) is a quantized equation for the gauge field, 
and therefore as stressed before, 
the Maxwell equation knows in advance the first quantization procedure. 

\section{Four Component Spinor}
Now, we derive the kinetic energy term of the fermion Lagrangian density. 
First, we assume that the Dirac fermion field should be written by four component 
spinor of eq.(1.3). This is based on the observation that electron has spin degree 
of freedom which is two. In addition, there must be positive and negative energy states 
since it is a relativistic field, and therefore 
the fermion field should have 4 components. This is the only ansatz which is set 
by hand in the derivation of the Dirac equation. 

\subsection{Matrix Elements}
The matrix elements 
$ \psi^\dagger \hat{O} \psi $ 
can be classified into 16 independent Lorentz invariant components as \cite{bd}
$$ \bar{\psi} \psi : {\rm scalar}, \ \ \ \ \ \ \bar{\psi}\gamma_5 \psi : 
{\rm pseudo-scalar} \eqno{ (3.1a)} $$
$$ \bar{\psi}\gamma_\mu \psi : {\rm 4 \  component \  vector }\eqno{ (3.1b)}  $$
$$  \bar{\psi}\gamma_\mu\gamma_5 \psi 
: {\rm 4 \  component \  axial-vector } \eqno{ (3.1c)} $$
$$ \bar{\psi} \sigma_{\mu \nu} \psi, : {\rm  6 \  component \ tensor} 
\eqno{ (3.1d)} $$
where $ \bar{\psi}$ is defined as 
$ \bar{\psi}  = \psi^\dagger \gamma_0 $. 
These properties are determined by mathematics, and therefore the vector 
current representation has nothing to do with physics. 

\subsection{Shape of Vector Current}
From the invariance consideration, one finds that the vector current $j_\mu$ 
must be written as
$$ j_\mu =C_0 \bar{\psi}\gamma_\mu \psi  \eqno{ (3.2)} $$
where $C_0$ is an arbitrary constant. 
Since we can renormalize the constant $C_0$ into the coupling constant $g$, we can set 
$$ C_0=1 . \eqno{ (3.3)} $$ 

\section{Shape of Lagrangian Density}
Now, we make use of the local gauge invariance of the Lagrangian density, and 
we assume the following shape that may keep the local gauge invariance
$$  {\cal L} = C_1\bar{\psi}\partial_\mu \gamma^\mu \psi-  
g\bar{\psi}\gamma_\mu \psi A^\mu   -{1\over 4} F_{\mu\nu} F^{\mu\nu}   \eqno{ (4.1)} $$
where $C_1$ is a constant. 

Now, we require that the Lagrangian density should be invariant under the local gauge 
transformation
$$ A_\mu \rightarrow A_\mu +\partial_\mu \chi  \eqno{ (4.2a)}  $$
$$ \psi \rightarrow e^{-ig\chi} \psi .  \eqno{ (4.2b)}  $$
In this case, it is easy to find that the constant $C_1$ must be 
$$ C_1=i . \eqno{ (4.3)} $$
Here, the constant $\hbar$ should be included implicitly into the constant $C_1$. 
The determination of $\hbar$ can be done only when one compares calculated results 
with experiment such as the spectrum of hydrogen atom.  

The Lagrangian density of eq.(4.1) still lacks the mass term. Since the mass term 
must be a Lorentz scalar, it should be described as
$$ C_2 \bar{\psi} \psi   \eqno{ (4.4)} $$
which is gauge invariant as well. 
This constant $C_2$ should be determined again by comparing the equation with experiment. 
By denoting $C_2$ as $(-m)$, we arrive at the Lagrangian density of a relativistic fermion 
which couples with the electromagnetic fields $A_\mu $
$$  {\cal L} = i\bar{\psi}\partial_\mu \gamma^\mu \psi-  
g\bar{\psi}\gamma_\mu \psi A^\mu -m\bar{\psi} \psi 
  -{1\over 4} F_{\mu\nu} F^{\mu\nu} .  \eqno{ (4.5)} $$
This is just the Lagrangian density for Dirac field and gives the Dirac equation. 

It is important to note that, in the procedure of deriving the Lagrangian density 
of eq.(4.5), we have not made use of the quantization condition of eq.(1.4). 
Instead, the first quantization is automatically done by the local gauge condition since 
the Maxwell equation knows the first quantization in advance.  

\section{Two Component Spinor}
The present derivation of the Dirac equation shows that the current density 
that can couple to the gauge field ${A}^\mu$ must be rather limited. Here, we discuss 
a possibility of finding field equation for the two component spinor \cite{gross}. 

When the field has only two components, 
$$ \phi = \pmatrix{ \phi_1 \cr 
 \phi_2 \cr  }  \eqno{ (5.1)} $$                  
then one can prove that 
one cannot make the current $j_\mu$ that couples with the gauge field  $A_\mu$. 
This can be easily seen since the matrix elements 
$ \phi^\dagger \hat{O} \phi   $ 
can be classified into 4 independent variables as
$$ {\phi}^\dagger \phi : {\rm scalar}, \ \ \ \ \ \ 
{\phi}^\dagger \sigma_k \phi: \ \ \ {\rm 3 \ component \  vector} . \eqno{ (5.2)} $$
Therefore, there is no chance to make four vector currents which may couple 
to the gauge field $A_\mu$. 

This way of making the Lagrangian density indicates that it should be difficult 
to find a Lagrangian density of relativistic bosons which should have two components. 

\section{Some results from the new picture}
In the present derivation of the Dirac equation, there is no need of the first 
quantization. Instead, the gauge principle 
is taken to be the most important guiding principle. At least, the first quantization 
does not have to be taken as the fundamental principle from which one can derive 
quantum mechanical equations. Here, one can derive the correspondence 
between ($E, \bm{p}$) and the differential operators 
($i\hbar {\partial\over{\partial t}}, \ -i\hbar\bm{\nabla} $)  as a consequence 
from the Dirac equation
$$ {\rm Dirac \ equation} \ \Rightarrow \ \ \ \  {
E}\rightarrow \ i\hbar {\partial\over{\partial t}},\ \ \ \ \ 
 {\bm{p}} \rightarrow \  -i\hbar\bm{\nabla} . $$
However, this replacement is not a fundamental principle any more as we saw above. 

Now, we may find some results from the new picture of quantum mechanics.  
Below we discuss some simple but interesting predictions from the new concept. 

\subsection{ No Klein-Gordon Equation}

The procedure of obtaining the Schr\"odinger equation in terms of 
the replacement of the momentum and energy by the differential operators 
in the classical Hamiltonian can be justified since the Dirac equation 
can be reduced to the Schr\"odinger equation in the non-relativistic limit. 
However, one cannot apply the relation eq.(1.4) to the energy 
dispersion relation with relativistic 
kinematics  since eq.(1.4) is not the fundamental principle any more. 

Historically, the Klein-Gordon equation was obtained by making use of the squared 
relativistic dispersion relation
$$ E^2=\bm{p}^2+m^2    \eqno{ (6.1)}  $$
and one replaced the energy and momentum by eq.(1.4). This leads to the following 
equation 
$$ E^2=\bm{p}^2+m^2  \Rightarrow \  -\hbar^2 {\partial^2\psi\over{\partial t^2}} 
= \left(-{\hbar^2}\bm{\nabla}^2+m^2\right) \psi  \eqno{ (6.2)}  $$
which is just the Klein-Gordon equation. However, this is only justified when the 
replacement of eq.(1.4) is the fundamental principle. If eq.(1.4) is only derived 
from the Dirac equation as a consequence of the field equation, then the Klein-Gordon 
equation cannot be justified any more. 
In other words, there is no way to derive the Klein-Gordon equation from any where, 
or at least, the Klein-Gordon equation has lost its foundation since the first 
quantization condition is not the fundamental principle. 
This is quite reasonable since there is no physical meaning of eq.(6.1) and 
the Klein-Gordon equation is just derived in analogy to the derivation 
of the Schr\"odinger equation. 

In addition, in spite of the fact that 
the Klein-Gordon equation is obtained by analogy to the non-relativistic 
dispersion relation, the Klein-Gordon field does not have any corresponding 
field in the non-relativistic limit. This peculiar behavior of the Klein-Gordon 
field is known, but it has never been discussed seriously. 
In this respect, the non-existence of the Klein-Gordon field must be quite 
natural even though bosons should, of course, exist  as composite objects of 
fermion and anti-fermion bound states \cite{okf}. 

\subsection{Incorrect Quantization in Polar Coordinates }

In quantum mechanics, one learns that one should not quantize  the classical Hamiltonian 
in polar coordinates. In order to see it more explicitly, one writes a free particle 
Hamiltonian in classical mechanics with its mass $m$
$$ H= {{p_r}^2\over 2m} +{{p_\theta}^2\over 2mr^2} 
+{{p_\varphi}^2\over 2mr^2\sin^2 \theta}  \eqno{ (6.3)}  $$
where $ p_r, \ p_\theta,\ p_\varphi $ are the generalized momenta in polar coordinates. 
In this case, if one quantizes the free particle Hamiltonian in classical mechanics 
with the quantization conditions 
$$ [r, p_r]=i\hbar,\ \ \ [\theta, p_\theta]=i\hbar, 
\ \ \ [\varphi, p_\varphi]=i\hbar   \eqno{ (6.4)}     $$
then one obtains
$$ H= -{\hbar^2\over 2m}\left({\partial^2\over{\partial r^2}}  
+{1\over r^2}{\partial^2\over{\partial \theta^2}}
+{1\over r^2\sin^2 \theta}{\partial^2\over{\partial \varphi^2}} 
\right) .  \eqno{ (6.5)}  $$
As is well known, this is not a correct Hamiltonian for a free particle 
in the polar coordinates, and the correct Hamiltonian for a free particle is, 
of course, given as
$$ H= -{\hbar^2\over 2m}\left({1\over r^2} {\partial\over{\partial r}}r^2
{\partial\over{\partial r}} \right) $$ 
$$ -{\hbar^2\over 2mr^2}\left( {1\over{\sin \theta}} {\partial\over{\partial \theta}}
\sin \theta {\partial\over{\partial \theta}} 
+{1\over \sin^2 \theta}{\partial^2\over{\partial \varphi^2}} 
\right)   \eqno{ (6.6)}  $$
which is obtained by transforming the Schr\"odinger equation in Cartesian coordinate 
into the polar coordinate.  However, one does not understand any reasons 
why one cannot quantize the Hamiltonian in  the polar coordinates 
as long as one starts from the canonical formalism with the canonical quantization. 
In quantum mechanics textbooks, one learns as an empirical fact that the transformation 
of the Cartesian to polar coordinates should be done after the quantization 
in the Cartesian coordinates is made in the Hamiltonian. 
  
Now, one can understand the reason why the quantization condition is valid 
for the Cartesian coordinates, but not for the polar coordinates. That is, 
in the present picture, the differential operators first appear in the Dirac equation 
and then one identifies the momentum as the corresponding differential operator.  
Therefore, one can obtain the Schr\"odinger equation from the Dirac equation, but not 
from the Hamiltonian of the classical mechanics by replacing the generalized momentum 
by the corresponding differential operator. In this sense, the canonical formalism 
of the classical mechanics is mathematically interesting, but it is not useful 
for quantum mechanics. 


\section{Discussions}

The quantization condition of $[x_i,p_j]=i\hbar\delta_{ij} $ is always the starting point 
of quantum mechanics. One learns in the textbooks that this condition is probably more 
fundamental than the momentum operator replacement of 
$ p_k\rightarrow -i\hbar{\partial \over{\partial x_k}}$. However, one cannot 
explain why the quantization condition is postulated, apart from the fact 
that the Schr\"odinger equation can be obtained from the Hamiltonian of classical 
mechanics. 

In this Letter, we have presented a new interpretation of the first quantization. 
Instead of the normal quantization, we have shown that the Lagrangian density 
of the Dirac field can be obtained from the local gauge invariance and 
the Maxwell equation. The first quantization procedure can be obtained from the Dirac 
equation as the consequence of the identification of the energy and momentum 
by the differential operators. 

In the historical derivation of the Schr\"odinger equation, one started from 
the Newton equation or the classical Hamiltonian dynamics. In this case, one had 
to assume the quantization condition of $ \bm{p}\rightarrow  -i\hbar \bm{\nabla}$. 
Here, we claim that the Dirac equation can be first obtained, and from the Dirac equation, 
one can derive the  Schr\"odinger equation by the Foldy-Wouthuysen-Tani transformation 
in the non-relativistic limit. From the Schr\"odinger equation, one can 
derive the Newton equation when one takes the expectation value, which is 
the Ehrenfest theorem. 

The present work is essentially based on the belief that the Maxwell equation 
and the local gauge invariance are the most fundamental principle. Once this 
standpoint is accepted, it is straightforward to derive the Dirac equation 
without the first quantization ansatz. 
Therefore, this procedure of first deriving the Dirac equation before 
the Schr\"odinger equation may present a new conceptual development 
in quantum mechanics, and one may learn  novel aspects in quantum mechanics 
in a future study.

\end{document}